\documentclass[lettersize,journal]{IEEEtaes}
\usepackage{color,array,amsthm}
\usepackage{graphicx,textcomp,stfloats}
\usepackage[caption=false,font=normalsize,labelfont=sf,textfont=sf]{subfig}

\jvol{XX}
\jnum{XX}
\jmonth{XXXXX}
\paper{1234567}
\pubyear{2022}
\doiinfo{TAES.2022.Doi Number}

\usepackage{amsmath,amsfonts}
\usepackage{bbm} %Added this for double-struck fonts
\let\labelindent\relax %needed to suppress error about legacy fxn when importing enumitem
\usepackage[shortlabels]{enumitem} %Added this for enumerating lists w/ letters
\usepackage{ctable} % for \specialrule command
\usepackage{float} %for more precise control of figure positions
\begin{document}
\title{Differentiable Rendering for \\ Synthetic Aperture Radar Imagery}
% \author{IEEE Publication Technology,~\IEEEmembership{Staff,~IEEE,} % <-this % stops a space
\author{Michael C. Wilmanski}%\member{Fellow, IEEE}
\affil{KBR National Security Technologies Group \\University of Texas at Austin}

\author{Jonathan I. Tamir, Member, IEEE}
\affil{University of Texas at Austin} 

% \author{THIRD C. AUTHOR Jr.}\member{Member, IEEE}
% \affil{University of Colorado, Boulder, CO 80309, USA}

%% \author{FOURTH D. AUTHOR}
%% \affil{University of Colorado, Colorado, USA}

\receiveddate{Manuscript received October 1, 2022; revised Februrary 24, 2023 and May 26, 2023; accepted July 25, 2023.\\
This work was a collaboration between the University of Texas at Austin and KBR, and was partially supported by KBR's office for Internal Research and Development.}
%This paragraph of the first footnote will contain the date on which you submitted your paper for review, which is populated by IEEE. It is IEEE style to display support information, including sponsor and financial support acknowledgment, here and not in an acknowledgment section at the end of the article. For example, ``This work was supported in part by the U.S. Department of Commerce under Grant BS123456.'' }
%% \accepteddate{XXXXX XX XXXX}
%% \publisheddate{XXXXX XX XXXX}

%\corresp{The name /of the corresponding author appears after the financial information, e.g. {\itshape (Corresponding author: M. Smith)}. Here you may also indicate if authors contributed equally or if there are co-first authors.}/
\corresp{{\itshape (Corresponding author: M. Wilmanski)}}

\authoraddress{
%The next few paragraphs should contain the authors' current affiliations, including current address and e-mail. For example, First A. Author is with the National Institute of Standards and Technology, Boulder, CO 80305 USA (e-mail: \href{mailto:author@boulder.nist.gov}{author@boulder.nist.gov}). 
Michael C. Wilmanski was previously part of the Computational Sensing and Imaging Lab at the University of Texas at Austin (Austin, TX 78712 USA) and currently works within the Sensors and Analysis directorate (Ann Arbor, MI 48108 USA) of KBR's National Security Technologies Group (Houston, TX 77005 USA) (e-mail: \href{mailto:wilmanski@utexas.edu}{wilmanski@utexas.edu}, \href{mailto:michael.wilmanski@us.kbr.com}{michael.wilmanski@us.kbr.com}).
%Second B. Author, Jr., was with Rice University, Houston, TX 77005 USA. He is now with the Department of Physics, Colorado State University, Fort Collins, CO 80523 USA (e-mail: \href{mailto:author@lamar.colostate.edu}{author@lamar.colostate.edu}). 
%Third C. Author is with the Electrical Engineering Department, University of Colorado, Boulder, CO 80309 USA, on leave from the National Research Institute for Metals, Tsukuba 305-0047, Japan (e-mail: \href{mailto:author@nrim.go.jp}{author@nrim.go.jp}).
Jonathan I. Tamir is with the Chandra Family Department of Electrical and Computer Engineering at the University of Texas at Austin (Austin, TX 78712 USA) and serves as faculty advisor of the Computational Sensing and Imaging Lab (e-mail: \href{mailto:jtamir@utexas.edu}{jtamir@utexas.edu}). }

%\editor{Mentions of supplemental materials and animal/human rights statements can be included here.}
\supplementary{Color versions of one or more of the figures in this article are available online at \href{http://ieeexplore.ieee.org}{http://ieeexplore.ieee.org}.}

\markboth{M. WILMANSKI ET AL.}{DIFFERENTIABLE RENDERING FOR SAR}
\IEEEoverridecommandlockouts
\IEEEpubid{\makebox[\columnwidth]{978-1-5386-5541-2/18/\$31.00~\copyright2018 IEEE \hfill}
\hspace{\columnsep}\makebox[\columnwidth]{ }}
\maketitle
\IEEEpubidadjcol

\begin{abstract}
    There is rising interest in differentiable rendering, which allows explicitly modeling geometric priors and constraints in optimization pipelines using first-order methods such as backpropagation. 
    Incorporating such domain knowledge can lead to deep neural networks that are trained more robustly and with limited data, as well as the capability to solve ill-posed inverse problems.  
    Existing efforts in differentiable rendering have focused on imagery from electro-optical sensors, particularly conventional RGB-imagery. 
    In this work, we propose an approach for differentiable rendering of Synthetic Aperture Radar (SAR) imagery, which combines methods from 3D computer graphics with neural rendering. 
    We demonstrate the approach on the inverse graphics problem of 3D Object Reconstruction from limited SAR imagery using high-fidelity simulated SAR data.
\end{abstract}


\section{INTRODUCTION}\label{sec:Introduction} 
    % \IEEEPARstart{T}{his} 
    S{\scshape ynthetic} aperture radar (SAR) is an imaging modality based on pulse-doppler radar. As a radar antenna moves, successive pulses of radio waves are transmitted to illuminate a scene. Echoed pulses from multiple antenna positions can be combined, which forms the synthetic aperture that enables higher resolution images than would otherwise be possible for a physical antenna \cite{IntroToRadar98}. In aerial and automotive vehicles, the vehicle's motion provides the movement needed to form the synthetic aperture.

    Recently, there is a rising interest in differentiable rendering for SAR imagery in order to perform downstream imaging tasks such as 3D object reconstruction. Similar to applications in electro-optical (EO) imagery, differentiable rendering is useful within the context of first-order optimization problems that may benefit from pixel-level supervisions flowing to 3D properties, including for training deep neural networks \cite{DRSurvey20}. While there have been successful demonstrations of methods for differentiable rendering of EO-domain imagery in recent years, these methods are not directly applicable to SAR-domain imagery, as SAR and EO images exhibit substantially different geometry and phenomenology. 
    Chief among these differences is the imaging wavelength. 
    As a result, many man-made objects will appear smooth at radar wavelengths and therefore be dominated by specular reflections. 
    For a similar reason, other objects (such as cloud cover) may appear transparent at radar wavelengths despite being opaque at visible light wavelengths.

    As SAR is a coherent imaging method and a form of active sensing, many priors about a scene are inherently known and controlled up to sensor and measurement error. Specifically, in this work we assume detailed knowledge about the scene's illumination as well as positions of the sensor and illuminator relative to the area on the ground being imaged. In addition to control over illumination and prior positional knowledge that can be expected, SAR imagery is invariant to many uncontrollable environmental effects that mar EO-domain imagery. This includes weather and atmospheric effects such as cloud cover, as well as illumination changes due to day/night cycles. Collectively, these differences mean that SAR-domain images have few scene parameters that are uncontrollable or unknown to the user, and thus are especially well-suited for differentiable rendering.

    \begin{figure*}[!t]
        \centering
        \includegraphics[width=0.90\textwidth]{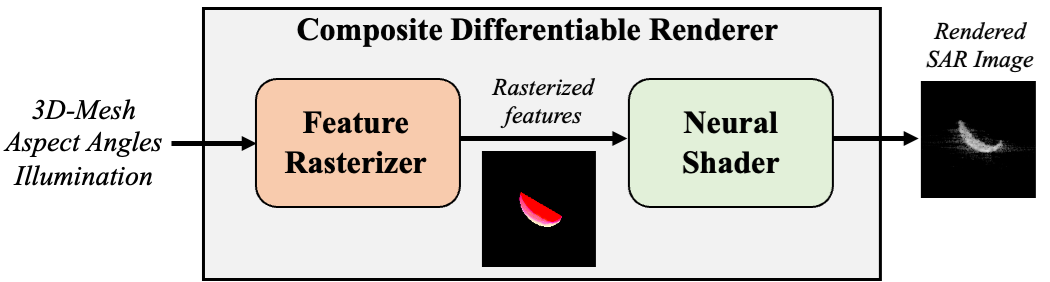}
        \caption{Block diagram of proposed composite differentiable rendering approach}
        \label{fig:compositeDiffRend}
    \end{figure*}
    
    Therefore, in this work we present a proof-of-concept and starting point for differentiable rendering in SAR imagery. Our proposed approach is composed of a Feature Rasterizer followed by a Neural Shader, depicted in the block diagram of Figure~\ref{fig:compositeDiffRend}. The Feature Rasterizer produces image-like feature maps that convey context about the structure and illumination of the scene. These features act as the input condition to the Neural Shader, which is implemented as a conditional generative adversarial network (CGAN) inspired by \cite{pix2pix17}. The Neural Shader processes the feature maps to produce realistic SAR scattering effects to complete the rendering procedure. 

    The primary benefit of this approach is its differentiability with respect to 3D scene parameters. 
    Due to its basis in rasterization, the differentiable rendering scheme is also computationally efficient and capable of real-time rendering; this attribute is important given its intended use in optimization problems, which typically involve many thousands of iterations. 
    The principal drawbacks to using rasterization are that the rendering scheme will inherently be limited to a first-order (single-bounce) approximation, as well as being limited to producing non-polarized magnitude imagery.
    
    To demonstrate the utility of our framework, we implement 3D object reconstruction from multiple SAR images as it is a common and highly useful downstream task enabled by differentiable rendering. Consequently, differentiation is with respect to an object's mesh vertices. 
    
    Our primary contributions are the following: 
    \begin{enumerate}
        \item We develop a differentiable rendering pipeline for SAR imagery, composed of a Feature Rasterizer and Neural Shader. The former uses softened rasterization to project mesh-based scene representations into 2D feature maps tailored to SAR's imaging geometry, while the latter uses a CGAN to estimate SAR scattering effects from the feature maps.
        \item We show how to train the system end-to-end for SAR shading using high-fidelity simulated SAR data. 
        \item We demonstrate proof-of-principle use via the inverse graphics problem of 3D object reconstruction.
    \end{enumerate}
    While our prototype renderer relies on high-fidelity SAR simulation, the framework is general and can be applied to physical SAR systems in the future.

    The remainder of this work is organized as follows. In Section~\ref{sec:RelatedWork} we discuss related work from the literature. In Section \ref{sec:Data} we describe the generation of a simulated dataset used for the experiments. In Sections \ref{sec:FeatureRasterizer} and \ref{sec:NeuralShader} we detail the feature rasterizer and neural shader components of the differentiable renderer, respectively. In Section \ref{sec:ReconExperiment} we describe experiments used to test the effectiveness of the differentiable renderer at solving the inverse problem of 3D object reconstruction. In Section \ref{sec:Discussion} we discuss the methods current limitations, future work, and provide a comparison of imagery from the differentiable renderer to collected data. Lastly, Section \ref{sec:Conclusion} provides a conclusion of this work's findings. 
    
\section{RELATED WORK}\label{sec:RelatedWork}
    \subsection{SAR Simulators}\label{sec:SarSimulators}
        The survey of SAR simulators in \cite{sarSims95} distinguishes two main categories of SAR simulation systems: 
        \begin{enumerate}[(a)]
            \item  SAR \textit{image} simulators, which directly produce focused SAR images.
            \item  SAR \textit{raw signal} simulators, which simulate raw sensor measurements. 
        \end{enumerate}  
        Category (a) simulators include examples such as RaySAR \cite{RaySAR16}, CohRaS\textsuperscript{\textregistered} \cite{CohRaS14}, and SARviz \cite{SarViz06}. A comparison of these and assessment of their limitations is provided in \cite{SarSims15}. Of these, SARviz is the only simulator that uses rasterization, while the others involve a form of ray tracing. RaySAR is the only project that is open source, but also has the most limited shading effects, relying on Phong shading \cite{Phong75}.
        
        Simulators of category (b) are usually based on physical optics approximations \cite{RemoteSensing86} and enhanced further with the shooting and bouncing rays (SBR) method pioneered in \cite{SBR89}. Instead of producing images directly, these physics-based simulators simulate raw sensor measurements, which the user must process themselves to coherently form images. These methods are the most realistic, but also computationally intensive.
        Examples in this category include Xpatch \cite{Xpatch95,Xpatch00}, CASpatch \cite{CASpatch09}, FACETS \cite{FACETS10}, 
        SigmaHat \cite{SigmaHat15}, FEKO \cite{FekoSuite715}, POFACETS \cite{pofacets00}, and SARCASTIC \cite{SarcasticSim17}. 
        Most simulators in this category are closed-source commercial products and do not explicitly support differentiable optimization. 
        Of these, POFACETS and SARCASTIC are the only projects that are open source, but lack some advanced features such as SBR.
        
        Our proposed differentiable rendering approach can be categorized as a SAR \textit{image} simulator, though the data used to evaluate it was produced by a SAR \textit{raw signal} simulator. Importantly, the simulators above are not suitable for solving inverse problems as they are not easily differentiable, and are therefore limited to black-box optimization.

        \begin{figure*}[!t] %moved from data gen section so it would be top of 3rd page
            \centering
            \includegraphics[width=\textwidth]{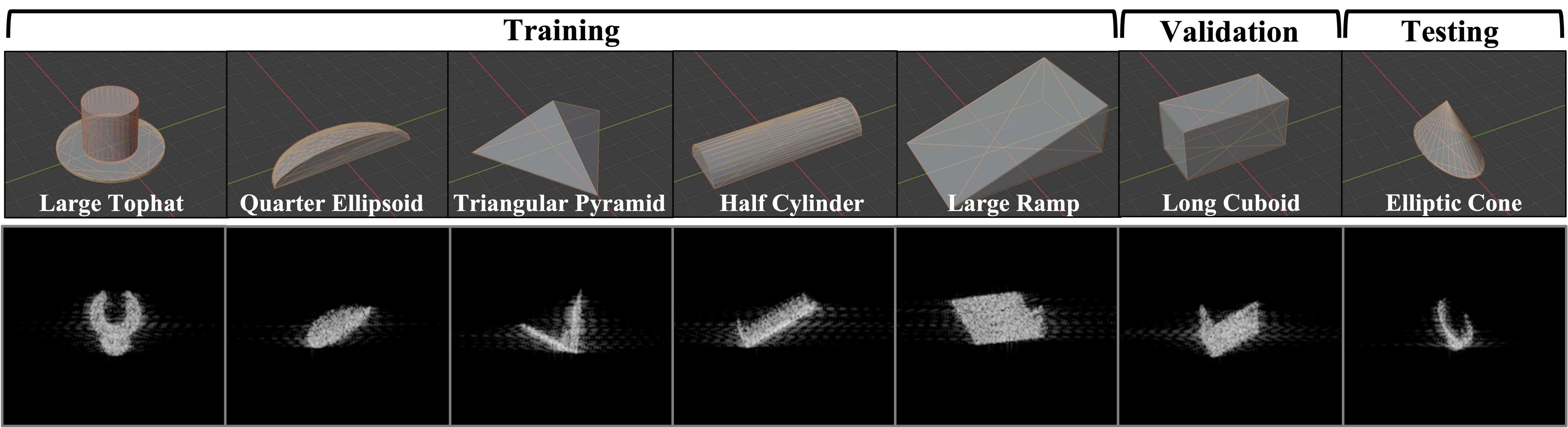}
            \caption{
            The seven objects used to generate train, validation, and test SAR images. (top) Ground-truth meshes viewed from Blender.
            (bottom) Fully processed simulated SAR images of corresponding objects at approximately similar aspect angles.}
            \label{fig:meshesOf7ObjectsAndSarImagesAfterProcessing}
        \end{figure*}
    
    \subsection{Neural and Differentiable Rendering}
        Neural rendering is an umbrella term referring to a broad field of methods that fuse machine learning models with techniques or domain-knowledge from computer-graphics. 
        A beneficial property of techniques in this family is that they are inherently differentiable; as they are composed of neural network building blocks, the resulting models can be easily backpropagated through.
        While they have been successful, it is worth noting that most neural rendering methods aim to manipulate existing images, rather than create images from raw scene representations \cite{NRSurvey20}.
        Such techniques are thus more akin to sophisticated pixel shaders, and not renderers in the formal sense. 
        The survey in \cite{NRSurvey21} describes these collectively as ``2D Neural Rendering'' methods, 
        and distinguishes them from an emerging paradigm it calls ``3D Neural Rendering.'' 
        
        Methods of this new paradigm, such as \cite{NeRF20}, learn to \textit{represent} a scene in 3D, using a differentiable rendering algorithm to facilitate training a neural network to represent the scene. 
        Methods of this branch of neural rendering are consequently best described as neural scene representations, and are yet another application of differentiable rendering.
        Cases such as \cite{NeRF20} use volume rendering techniques, which are naturally differentiable while also memory inefficient.
        
        Besides volumetric techniques, most differentiable rendering work in the literature has focused on rasterization-based methods from computer graphics \cite{DRSurvey20}. 
        Rasterization is computationally-efficient, but not naturally differentiable. However, smoothed variants  
        have emerged \cite{meshRenderer18,pix2vex19,softRasterizer19}, which have made rasterization-based methods viable. 
        Physics-based differentiable rendering methods which model global light transport effects exist as well \cite{DiffMCRT18,Mitsuba2}, but are substantially slower. These methods are the most accurate, but the hypothetical improvements in rendering accuracy may not be worth the increased computational cost. 
        
        Use of neural and differentiable rendering methods in the context of SAR has so far been limited.
        Several works have proposed using CGANs for image-to-image translation from SAR to EO imagery \cite{SarToOptical19,SarToOptical20,SarToOptical22}. 
        A more related example is \cite{embeddedDnnSarSim21}, which attempts to improve the realism of the RaySAR SAR image simulator \cite{RaySAR16} by training a deep neural network to estimate realistic SAR scattering effects.
        This falls under the ``2D Neural Rendering'' paradigm and is similar in principal to our use of a CGAN for the Neural Shader.
        
        Other instances of differentiable rendering for SAR include \cite{MSReport21}, which is an academic report of an earlier iteration of this work, 
        and very recently \cite{DiffSarRendAndRecon22}, which proposes a scheme for 3D object reconstruction from SAR images using differentiably-rendered silhouettes.
        Although the latter's proposed scheme is capable of outputting simulated SAR images in tandem with the silhouettes, the simulated SAR images do not support differentiation, and thus the scheme is not well-suited for inverse problems such as 3D object reconstruction.

\section{DATA GENERATION}\label{sec:Data}
    Computer-aided design models for a total of seven object shapes were created, pictured in Figure~\ref{fig:meshesOf7ObjectsAndSarImagesAfterProcessing}. The objects were created in Blender \cite{Blender21} and exported as triangulated meshes. The ground plane gridded lines are spaced in units of one meter, making these objects similar in size to medium or large vehicles.

    The first five objects were used to generate training data for the Neural Shader. The Long Cuboid was used to generate validation data to inform hyperparameter tuning for both the Feature Rasterizer and Neural Shader. The Elliptic Cone was reserved for generating test-only data and used only after all aspects of the end-to-end rendering process were frozen.
    
    \begin{figure}[!t]
        \centering
        \includegraphics[height=5.4cm]{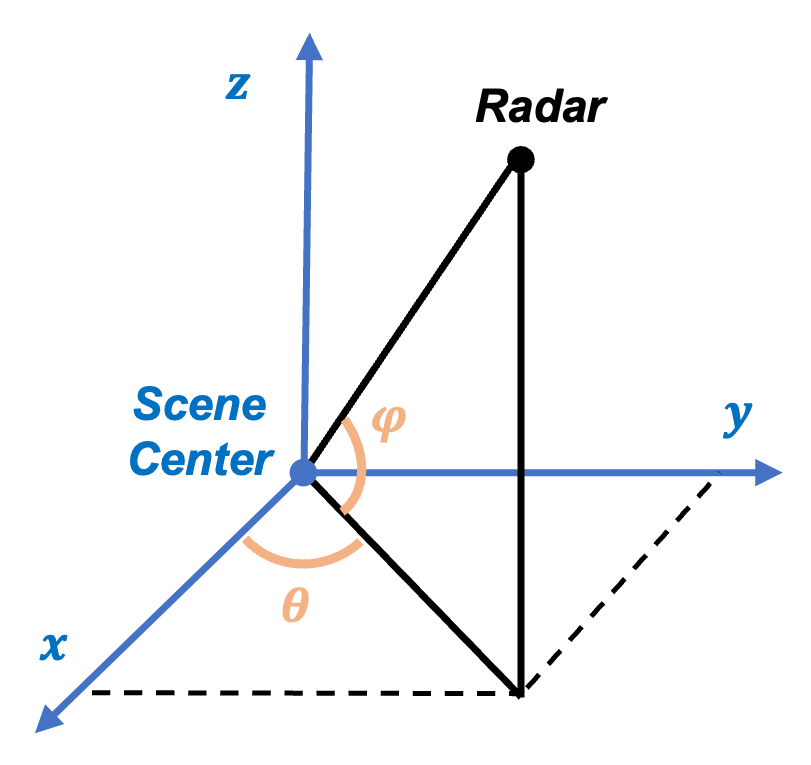}
        \caption{Collection geometry for aspect angles $\theta$ (azimuth) and $\phi$ (elevation).}
        \label{fig:geometryAspectAngles} 
    \end{figure}
    
    The data generated for this study consist of six distinct elevation angles, evenly spaced between [10\textdegree, 60\textdegree], and 36 distinct azimuth angles, evenly spaced between [0\textdegree, 350\textdegree]. This gives a total of 216 distinct aspect-angle combinations per object. A geometric representation of the aspect angles is illustrated in Figure~\ref{fig:geometryAspectAngles}. Data are assumed to be monostatic, meaning transmitter and receiver are co-located, which is the most common scenario for SAR.

    Every object is assumed to be made of perfect electric conducting material. This simplifies the simulations and is a reasonable assumption for most metallic objects, though it is straightforward to simulate other materials at a modest increase in data generation time. 
    The texture of the surfaces of each object are assumed to be of moderate roughness (\(\sim \)0.5cm RMS of surface variability). 
    Each simulation uses nominally 256 pulses at X-band frequencies, spanning 11\textdegree ~of synthetic aperture and having 2GHz of total bandwidth. The resulting SAR collections have a very fine spatial resolution of approximately 0.075m in both range and cross-range directions prior to tapering.
        
    The raw Fourier data (known as ``phase history'' in the SAR community) acquired from the simulations are first Taylor-weighted as described in \cite{BlueBook95}. Adding such a taper to the data is used primarily to decrease the prevalence of sidelobes in the formed images, albeit at the expense of slight degradation in spatial resolution. 
    Next, images are formed via Backprojection, based on \cite{ImgFormationToolbox10}. This implementation also projects the SAR images into the ground-plane.

    \begin{figure}[!t]
        \centering
        \includegraphics[height=3.5cm]{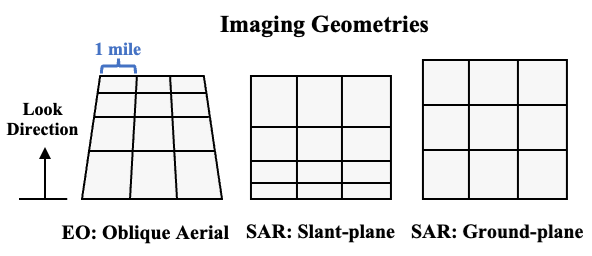}
        \caption{Comparison of imaging geometries in EO and SAR when imaging points on the ground.}
        \label{fig:imagingGeometries} 
    \end{figure}

    \begin{figure}[!t] 
        \centering
        \subfloat[]{\includegraphics[width=2.5in]{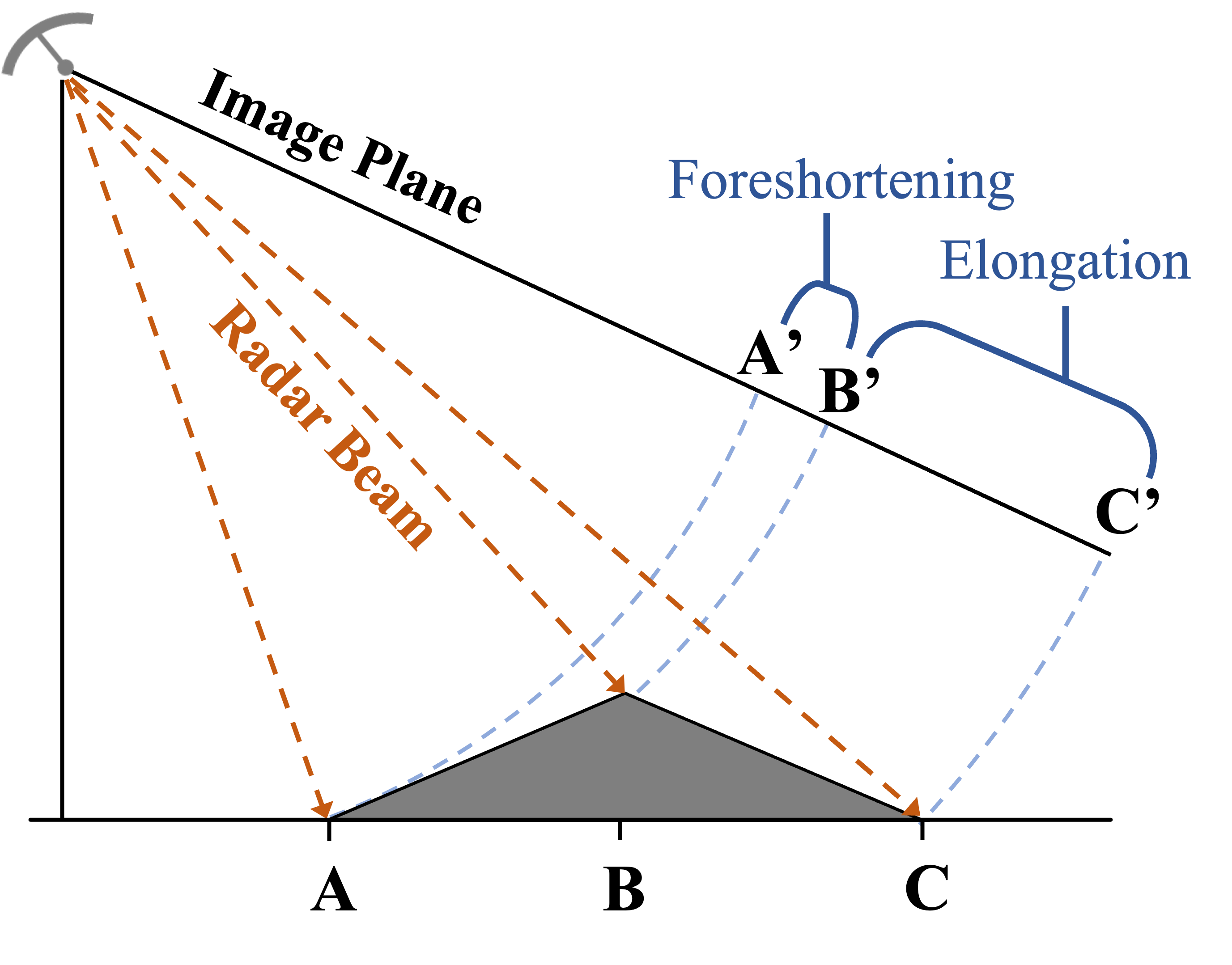}%
        \label{foreshorteningElongation}}
        \vfil %\hfil
        \subfloat[]{\includegraphics[width=2.5in]{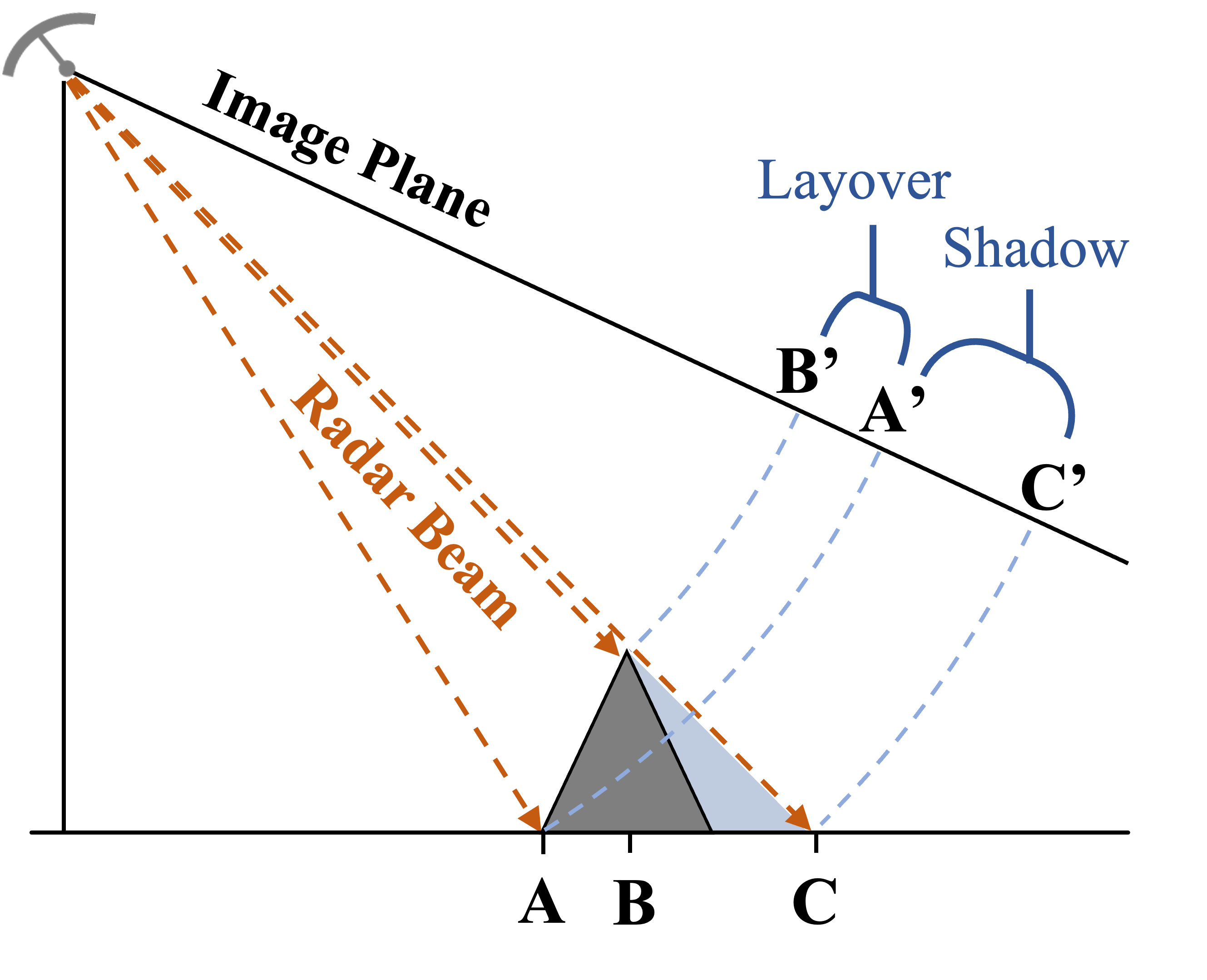}%
        \label{layoverShadow}}
        \caption{Graphical representations of (a) foreshortening and elongation and (b) layover and shadowing.}
        \label{fig:foreshorteningElongationLayoverShadow}
    \end{figure}

    Following image formation, the complex-valued images may optionally be augmented with one of two types of noise for additional training data diversity.
    The first type of noise is additive and follows a complex-normal distribution. 
    The absolute value of a zero-mean complex-normal random variable will be Rayleigh-distributed. For this reason, the Rayleigh distribution has traditionally been used to model random clutter in magnitude-detected SAR imagery \cite{SarClutterModeling04}.
    Consequently, pixels that had an original magnitude of zero will follow a Rayleigh distribution after augmentation and magnitude-detection.
    As the original magnitude of a pixel increases, its distribution after augmentation and magnitude-detection will shift toward being normally-distributed, centered around the original magnitude. 
    The second type of noise is multiplicative and follows a uniform distribution. This type of noise is typically used to model SAR speckle \cite{SarSpeckleNoise16}. 
    Equations for applying both types of noise augmentation are given as: 
    \begin{align}\label{eqn:AddNoise}
        \widetilde{S}_{ij} & = S_{ij} + Z_{ij} \; \\ \label{eqn:MulNoise}
        \widetilde{S}_{ij} & = S_{ij} \cdot (1+Y_{ij}),\; 
    \end{align}
    where $S$ and $\widetilde{S}$ represent the original and augmented signals, respectively. $Z$ represents zero-mean complex-normal noise, $Y$ represents zero-mean uniform noise, and subscripts $i$ and $j$ specify a pixel. During optimization, images are augmented at random, with either equation \ref{eqn:AddNoise} (25\%), equation \ref{eqn:MulNoise} (25\%), or neither (50\%).
    
    After image formation and optional augmentation, the complex-valued images are magnitude-detected and remapped via the Piecewise Extended Density Format (PEDF), as implemented in the SarPy library \cite{SarPy21}. 
    PEDF remaps pixel magnitudes to a log-like scale, to account for SAR's naturally high dynamic range and improve interpretability.
    Figure~\ref{fig:meshesOf7ObjectsAndSarImagesAfterProcessing} shows examples of resulting images after this remapping.

\section{SAR FEATURE RASTERIZER}\label{sec:FeatureRasterizer}
    In this section we describe the SAR Feature Rasterizer, which projects mesh-based scene representations into 2D feature maps that convey context about the structure and illumination of the scene.  
    The Feature Rasterizer consists of three stages: Coordinate Transformations, Rasterization, and Shading.
    Standard computer graphics rendering follows analogous stages; however, each stage requires alterations either to accommodate backpropagation, account for SAR's unique imaging geometry, or for the features we want to produce for the Neural Shader. The implementation of this Feature Rasterizer extends and is based on the PyTorch3D API \cite{pytorch3D20}.

    \subsection{SAR Imaging Geometry and Coordinate Transformations}
        
        \newcolumntype{P}[1]{>{\centering\arraybackslash}p{#1}} %define a new column type with horizontal centering
        \setlength{\tabcolsep}{4pt}
        \begin{table*}[t]
            \centering
            \caption{Coordinate transformation details for SAR ground-plane geometry.}
            \label{table:coordinateTransforms}
            %\resizebox{\textwidth}{!}{ %wrapper to shrink the table proportionately to fit text width (allowed in the context of tables)
            \begin{tabular}{|P{17mm}|P{17mm}|P{17mm}|P{105mm}|}
                \hline %\noalign{\smallskip}
                \textbf{Transform Name} & \textbf{Input \mbox{Coordinates}} & \textbf{Output \mbox{Coordinates}} & \textbf{Description}\\
                %\noalign{\smallskip}
                \specialrule{.1em}{.05em}{.05em} %special thicker line 
                % \noalign{\smallskip}
                View Transform & ~~World\newline Space & ~~View\newline Space & Coordinates relative to camera's point of view\\
                \hline
                Range Transform & ~~~View\newline Space & ~~Range\newline Space & z-coordinates (depth, or distance from camera) become new y-coordinates\\
                \hline
                \mbox{Ground-plane} Projection & ~~~Range\newline Space & ~~~Ground\newline Space & Stretches perceived y-dimension s.t. each pixel maps to a consistent-sized cell on ground\\
                \hline
                Orthographic Projection & ~~~Ground\newline Space & ~~~NDC\newline Space & Normalizes coordinate values into range natively interpretable by graphics processor (skips perspective projection, as distance from camera does not affect perceived size) \\
                \hline
            \end{tabular}%}
        \end{table*}
        \setlength{\tabcolsep}{1.4pt}

        \begin{figure*}[t]
            \centering
            \includegraphics[width=0.98\textwidth]{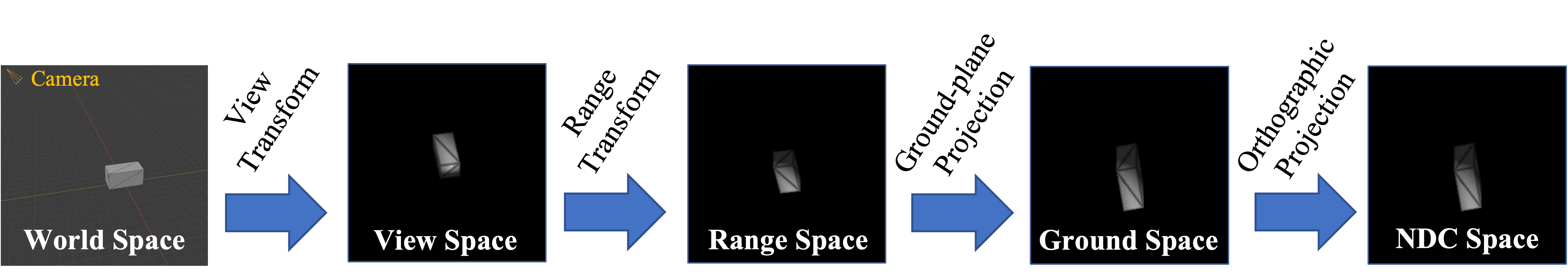}%coordinateTransformSeq.png}
            \caption{Coordinate transform sequence for SAR ground-plane imaging geometry.}
            \label{fig:coordinateTransformSeq}
        \end{figure*}
    
        SAR's unique imaging geometry makes adapting methods in differentiable rendering designed for EO imagery non-trivial. 
        For rasterization-based methods, imaging geometry is controlled primarily through the coordinate transformation sequence.
        Consequently, understanding the geometry differences between these imaging modalities is necessary to design a coordinate transformation sequence. 

        In EO imagery, distance from the sensor affects an object's perceived size. This results in the ``oblique aerial'' imaging geometry, seen in Figure~\ref{fig:imagingGeometries}. In computer graphics, this effect is known as ``perspective,'' and is typically accounted for via a ``perspective projection'' during the conversion to Normalized Device Coordinates (NDC) \cite{learnOpenGL20}. 
        However in SAR imagery, distance from the sensor determines which range-bin reflected energy funnels into, and thus affects an object’s perceived \textit{position} in range. 

        This leads to a natural ``slant-plane'' imaging geometry for points on the ground. SAR images may be projected into the ``ground-plane'' as part of image formation, nominally mapping image pixels to equally spaced points along the ground.
        Distance to the sensor is also influenced by an object's height. Consequently, imaging objects of variable height will yield sampling distortions known as ``foreshortening,'' ``elongation,'' and ``layover,'' as well as SAR's signature self-shadowing effect. These phenomena are described in detail by \cite{BlueBook95,ImgVidHandbook00} and graphically depicted in Figure \ref{fig:foreshorteningElongationLayoverShadow}.
        
        The goal of the coordinate transformation stage is to transform mesh coordinates appropriately to get the desired SAR ground-plane imaging geometry, including foreshortening/elongation/layover effects. Coordinate Transformation sequences must start in world space coordinates and end in NDC. The primary sequence is described in Table~\ref{table:coordinateTransforms} and illustrated in Figure~\ref{fig:coordinateTransformSeq}. If the ground-plane projection were to be omitted, the result would be a SAR slant-plane imaging geometry instead. An equivalent EO ``oblique aerial'' imaging geometry would require omitting both the range transform and ground-plane projection, as well as changing the orthographic projection to a perspective projection.
        
    \subsection{Rasterizing and Shading Feature Maps}

        \begin{figure}[t]
            \centering
            \includegraphics[width=0.35\textwidth]{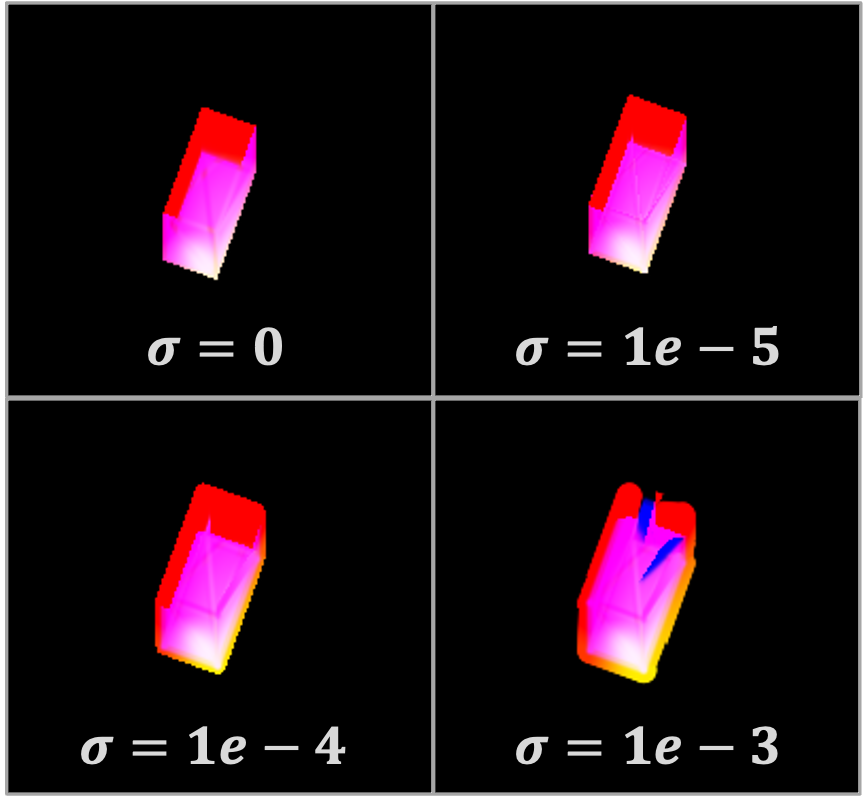}
            \caption{Visual artifacts from soft rasterization using varying levels of blending radius $\sigma$.}
            \label{fig:softRasterizationArtifacts}
        \end{figure}

        The rasterization method is based on the soft rasterizer of \cite{softRasterizer19}. The method in \cite{softRasterizer19} blends facet contributions for pixels near edges to mitigate the discontinuities that would arise if trying to backpropagate through a standard rasterization operator. Soft rasterization will yield slightly distorted images but allows backpropagating more useful gradients. Examples of artifacts from soft rasterization are shown in Figure~\ref{fig:softRasterizationArtifacts}.
        
        Rasterizing with our custom coordinate transform sequence produces the fragments necessary to shade the first two feature maps. The first feature map is a basic silhouette of the object. This can be produced by making a mask out of fragments that have no corresponding facets. The second feature we want is pixel-wise surface-normal information, relative to the direction of the sensor and illumination. For monostatic imagery, this is given as:
        \begin{align}
            d & = - \hat{q} \cdot \hat{n} \,,
        \end{align}
        where $\hat{q}$ is the direction of the illumination, $\hat{n}$ is the per-pixel surface-normal of the object, and $\cdot$ denotes the inner-product.
        
\begin{figure}[t]
            \centering
            \includegraphics[width=0.35\textwidth]{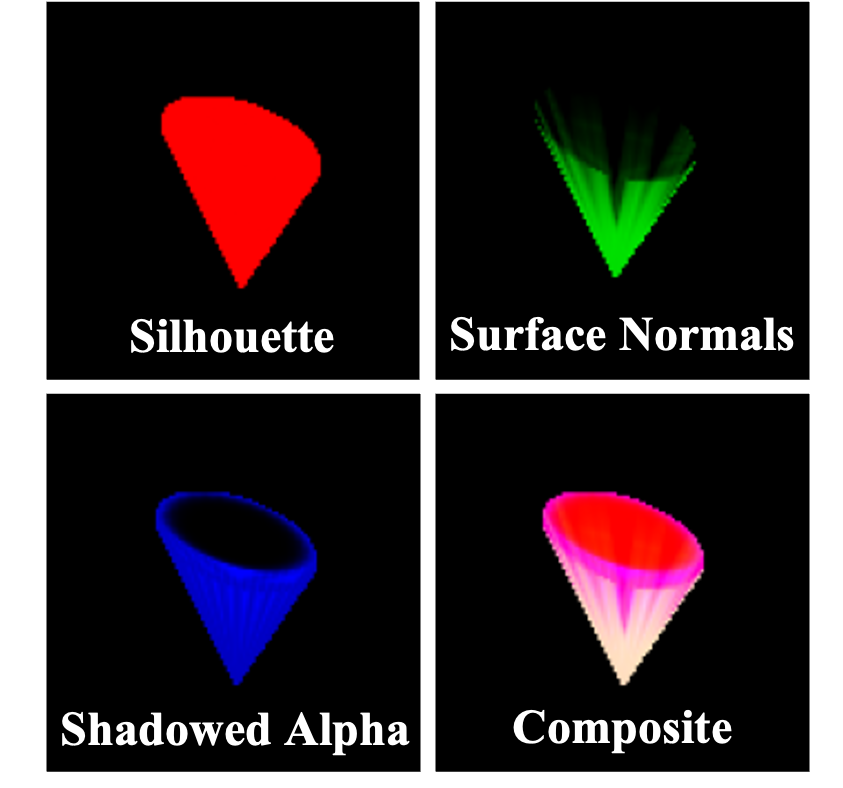}
            \caption{Shaded feature maps.} 
            \label{fig:shadedFeatureMaps}
        \end{figure}

        The last feature map to shade is an alpha channel. 
        Deriving the alpha channel is complicated by SAR's self-shadowing effect. 
        To address this, we rasterize a reference view from the perspective of the radar, which is used to determine which facets should be visible in its main beam. 
        Any facet that is not visible from this reference viewpoint is masked out from contributing in the main view when calculating the alpha channel feature map.
        An example of all three shaded feature maps is shown in Figure~\ref{fig:shadedFeatureMaps}. 

\section{NEURAL SHADER}\label{sec:NeuralShader}
    The Neural Shader is trained to predict realistic SAR shading/scattering effects conditioned on the features output by the Feature Rasterizer. Given this is effectively a paired image-to-image translation problem, we choose a model based on Pix2Pix \cite{pix2pix17}. While some newer approaches have outperformed Pix2Pix, we choose it as a basis for its simplicity in this proof of principle work. The most noteworthy deviation made is to use batch renormalization \cite{batchRenorm17} instead of instance normalization.
    Batch renormalization addresses the original issues that motivated the authors to use instance normalization, while improving consistency and accuracy.
    
    \subsection{Training}
        The feature map inputs used in training involve minor data augmentation via differing values for blending radius $\sigma$ (0, 1e-5, 1e-4, 1e-3), like those seen in Figure~\ref{fig:softRasterizationArtifacts}. This adds diversity to the training data and should help the CGAN be more robust to distortions caused by soft rasterization. As described in Section \ref{sec:Data}, the SAR images are augmented via additive (clutter-like) and multiplicative (speckle-like) types of noise. 
        
        The CGAN is trained for 200 epochs with early stopping based on performance with the validation subset. Both the generator and discriminator are trained with the AdamW optimizer \cite{AdamW18}. AdamW is a slight improvement over the original Adam optimizer \cite{adam15} by making a correction to the implementation of weight decay. Learning rates start at 2e-4 and decay linearly every epoch to reach 2e-5 by the end of training.
        
    \subsection{Results}\label{sec:NeuralShaderResults}
\begin{figure*}
            \centering
            \includegraphics[width=0.98\textwidth]{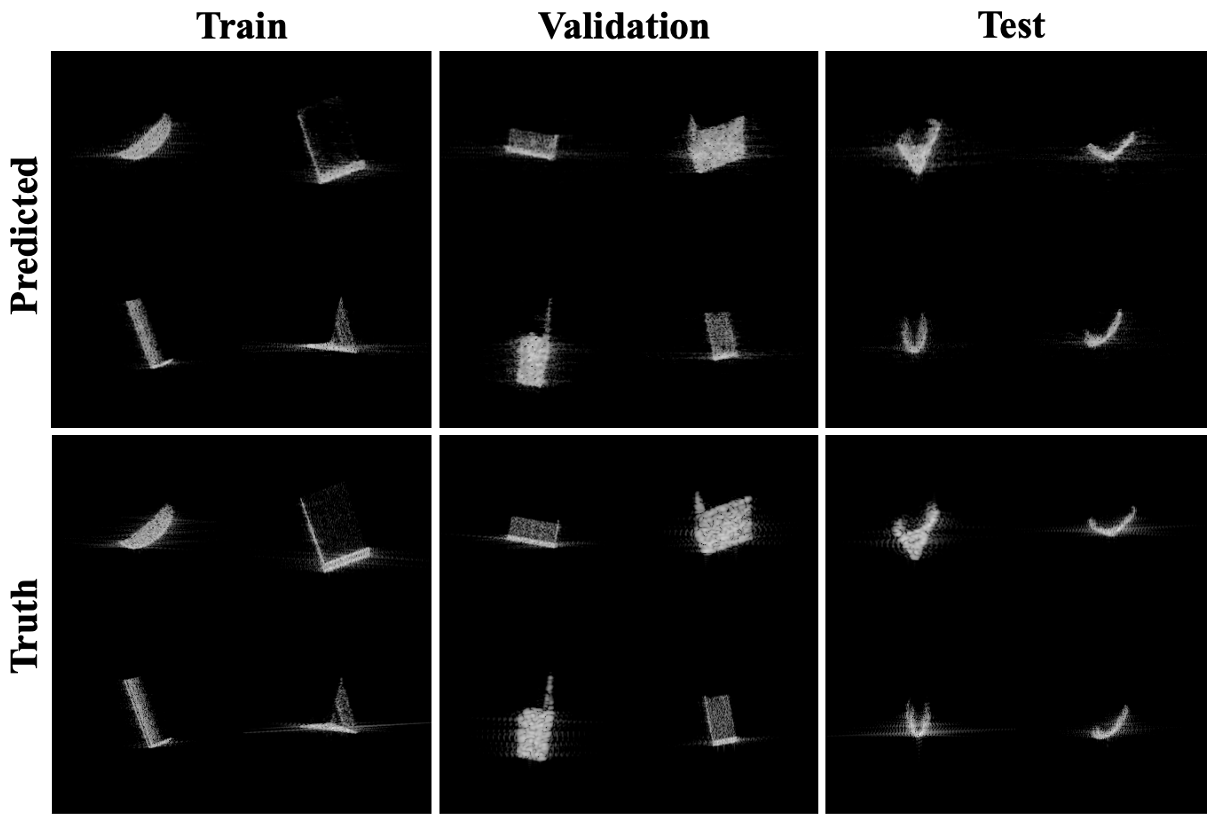}
            \caption{Example images generated by the fully trained CGAN.}
            \label{fig:neuralRenderingPreds}
        \end{figure*}

        Figure~\ref{fig:neuralRenderingPreds} shows example images generated by the fully trained CGAN. Although the most important results are those for the test object, we include predictions from objects in the validation and training sets for additional comparison.
        Empirically, the predicted and truth images appear very similar, with the most apparent difference being in the sidelobe/glint effects.
        
        The train, validation and test object sets are compared in terms of the loss produced when generating images of each with the CGAN, in Table~\ref{table:averageCGANLosses}. 
        The $L_{cGAN}$ and $L_{L1}$ losses in Table \ref{table:averageCGANLosses} are similar to those found in \cite{pix2pix17}, except that the weighting for $L_{cGAN}$ is decreased from 1 to 0.5. 
        $L_{cGAN}$ is a measure of how well the generator fools the discriminator, while $L_{L1}$ is the scaled L1 distance between predicted and truth images.
        
        We also include an energy-normalized metric, $L_{L2}^*$, to account for bias due to varying object sizes. The sum-squared-error of the residual is normalized by the energy of the label image, and this ratio is converted to a dB scale. The metric is defined in equation \ref{eqn:NormalizedMetric}, where norms are over pixels.
        \begin{align}\label{eqn:NormalizedMetric}
            L_{L2}^* & = 10 \log_{10} \left( \frac{\|y-\hat{y}\|^2}{\|y\|^2}\right) .
        \end{align}
        
        \setlength{\tabcolsep}{6pt}
        \begin{table}[t]
            \centering
            \caption{Average loss comparison by object set for trained CGAN.}
            \label{table:averageCGANLosses}
            \begin{tabular}{|c|c|c|c|c|}
                \hline %\noalign{\smallskip}
                \textbf{Data Set} & $\boldsymbol{L_{cGAN}}$ & $\boldsymbol{L_{L1}}$ & $\boldsymbol{L_{total}}$ & $\boldsymbol{L_{L2}^*}$\\
                %noalign{\smallskip}
                \specialrule{.1em}{.05em}{.05em} %special thicker line
                %\noalign{\smallskip}
                Train & 0.46 & 1.45 & 1.91 & -8.93 \\
                \hline
                Validation & 0.47 & 1.44 & 1.91 & -8.45 \\ 
                \hline
                Test & 0.49 & 0.94 & 1.43 & -6.38 \\ 
                \hline
            \end{tabular}
        \end{table}
        \setlength{\tabcolsep}{1.4pt}
        
        When using the energy-normalized metric $L_{L2}^*$, we can see that loss is lowest on the training set as expected, with an increase of +0.48 dB loss on the validation set, and an additional +2.07 dB loss on the test set. 
        This is the pattern of losses we would expect, 
        which suggests the CGAN suffers slightly from both overfitting to the training data and over-tuning to the validation data.

\section{3D RECONSTRUCTION EXPERIMENT}\label{sec:ReconExperiment}
    In this section we evaluate the proposed differentiable renderer in the task of object reconstruction. Specifically, we use the renderer to reconstruct a mesh of a new object, given only SAR images of the object at various known aspect angles.
    
    \subsection{Losses}
        This subsection details losses used for the 3D reconstruction experiments. 
        The total loss is aggregated from five sources, with weights that were tuned based on reconstruction performance using the validation object, and is given as:
        \begin{align}
            \label{eqn:L_total} %The # for this eqn runs onto a new line...
            L_{total} & = \mathbbm{E} \! \left[ |y \! - \! \hat{y}|^2 \right] + 1.5L_l + 0.02L_n + 0.03L_e + 0.4L_f .
        \end{align}
        
        The first term uses pixel-wise mean-squared-error to compare the high-fidelity simulated SAR images (at known aspect angles) to output predictions of the differentiable SAR renderer (at the same aspect angles). 
        Despite the notorious fickleness of SAR speckles, we found this loss still works reasonably well for demonstration purposes and is simple. 
        The other loss terms are for regularization of the mesh shape and are described next. The regularization terms are primarily intended to prevent the mesh from becoming tangled or trapped in a poor local minima while fitting.
        
        \subsubsection{Laplacian} 
            We use the ``uniform'' weighting of Laplacian mesh regularization based on \cite{laplacian99,laplacian06}, which discourages sharp vertex corners. If $v_i$ represents the $i$th vertex of the mesh, $S_i$ is the set of vertices it shares edges with, and $\mathbbm{E} [S_i]$ is its centroid, then the Laplacian regularization loss can be computed as: 
            \begin{align}
                L_{l} & = \sum_i \|\mathbbm{E} [ S_i ] - v_i\| \,. 
            \end{align}
        
        \subsubsection{Normal Consistency} 
            We use normal consistency mesh regularization as implemented in \cite{pytorch3D20}, which encourages adjacent faces to have similar surface-normal directions. If $\hat{n}_{a_k}$ and $\hat{n}_{b_k}$ are surface normal vectors of faces sharing the $k$th edge, then the normal consistency regularization loss is given as:
            \begin{align}
                L_{n} & = \sum_k 1 - \hat{n}_{a_k} \cdot \hat{n}_{b_k} \,. 
            \end{align}
        
        \subsubsection{Edge Length} 
            Edge length regularization is meant to encourage mesh edges to be of similar length. This has the effect of encouraging vertices to be uniformly distributed along the mesh's surface. 
            If $\gamma$ is the set of the mesh's edge lengths, the edge length regularization loss can be written as:
            \begin{align}
                L_{e} & = \sum_k \left( \mathbbm{E}[\gamma] - \gamma_k \right)^2 \!.
            \end{align}
        
        \subsubsection{Floor-Plane} 
            Since only overhead views are realizable, vertices on the underside of the mesh will be occluded from every view, and thus will not receive guiding gradient information from the Primary loss. 
            However, it is known as a prior that such vertices are likely to reside on the scene's floor plane, as opposed to being suspended in air.
            
\begin{figure}[t] %\begin{wrapfigure}{}{0.41\textwidth}
                \centering
                \includegraphics[height=5.0cm]{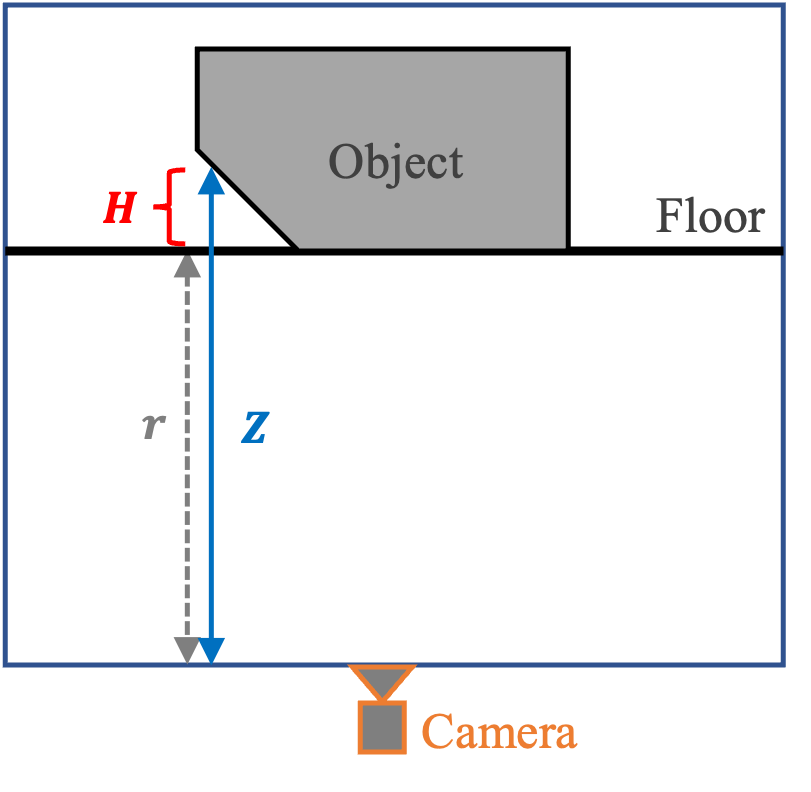}
                \vspace{-0.2cm} %brings captions slightly closer so graphic can be slightly taller
                \caption{Geometric representation of floor-plane regularization.}
                \label{fig:floorPlaneQuantities}% 
            \end{figure}

            We created the `floor-plane' regularizer to encourage this by first rasterizing the object from a bottom-up view. 
            The depth buffer ($Z$) from the raster fragments can be used to derive a 2D height map ($H$) of the underside of the mesh.
            We penalize the heights as:
            \begin{align}\label{eqn:floorPlaneHeight}
                H & = r - Z \\
                L_{f} & = \mathbbm{E}[H^2] \,,
            \end{align}
            where $r$ is the distance between the camera and the scene's floor-plane. It is assumed that the floor-plane has a height of zero. Figure~\ref{fig:floorPlaneQuantities} provides a geometric interpretation of the quantities in equation \ref{eqn:floorPlaneHeight}.

            %\end{wrapfigure}
    
    \subsection{Mesh Optimization Loop and Settings}
        The 3D object reconstruction is done in two levels, to reconstruct the mesh in a coarse-to-fine manner. Both levels use the stochastic gradient descent optimizer with a learning rate of 1.0, momentum of 0.9, and dampening of 0.9. Each level updates the mesh iteratively as follows:
        
        \begin{enumerate}
            \item A label SAR image of the target object and its aspect angles are chosen.
            \item The estimated mesh is rendered via our renderer at the same aspect angles. 
            \item The total (regularized) loss is computed by equation \ref{eqn:L_total}, using the prediction rendered by the model and the chosen label SAR image.
            \item The mesh is updated based on backpropagated gradients from the loss.
        \end{enumerate} 
        
\begin{figure}[t]%\begin{wrapfigure}{R}{0.42\textwidth}
            \centering
            \includegraphics[width=0.4\textwidth]{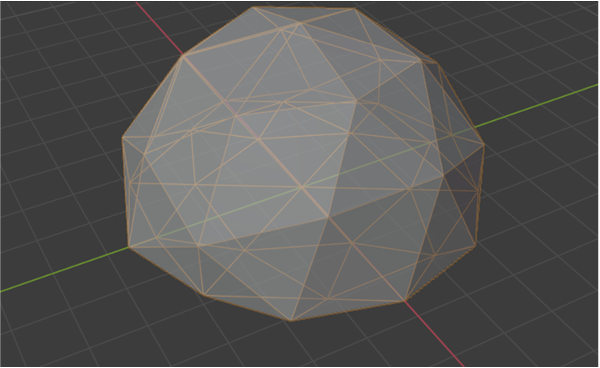}
            \caption{Initial dome mesh.}
            \label{fig:initialMesh}
        \end{figure}

        During the first level of mesh-fitting, the estimated mesh is initialized with a large dome shape of 42 vertices and 80 triangle faces, as seen in Figure~\ref{fig:initialMesh}. We choose a low vertex count to help the optimization loop avoid local minima early on. The soft rasterizer blending radius is $\sigma$=1e-3. The mesh at this level is updated for 1080 iterations using a batch size of two.

        %\end{wrapfigure}
        
        After the first level of mesh-fitting, the mesh estimate is up-sampled by inserting new vertices in the middle of each edge, resulting in a mesh with 162 vertices and 320 triangle faces. This up-sampled mesh initializes the second level. The soft rasterizer blending radius is also lowered to $\sigma$=2.5e-4. The mesh is updated for 540 iterations using a batch size of four. The larger batch size of the second level reduces variance of the gradients, helping the mesh fine-tune. 

    \subsection{Metrics for 3D Reconstruction}
        For quantitative evaluation, we consider two main metrics, as well as translation-invariant versions of each.
        The first metric is voxelized intersection over union (IoU), while the second is voxelized cosine similarity (CS). CS is a common comparison metric across many domains, although IoU is more common specifically for comparing 3D volumes. The metrics can be defined as:  
        \begin{align} \label{eqn:IoU}
            IoU & = \frac{\sum (\hat{V} \cap V)}{ \sum (\hat{V} \cup V)} \\ 
            CS & = \frac{<\hat{V}, V>}{ \| \hat{V} \|_2 \| V \|_2 } \,, \label{eqn:CS}
        \end{align}
        where $\hat{V}$ and $V$ are voxelized volume representations of the reconstructed and ground-truth meshes, respectively. Norms and summations are over voxels, and each voxel is boolean-valued to indicate whether it lies inside or outside the mesh.
        Note that since voxels are boolean-valued, the numerators of equations \ref{eqn:IoU} and \ref{eqn:CS} will be equal. 
        The boolean-valued voxels also mean that the denominator in equation \ref{eqn:CS} is equivalent to the geometric-mean of the two volume sizes, 
        while the denominator in equation \ref{eqn:IoU} is instead the size of the union between the two volumes. 
        Although both metrics yield values between [0,1], the difference in denominators makes IoU less-forgiving of small deviations between the compared volumes.
        
        Translation-invariant versions of these metrics may also be useful. 
        This removes the influence of potential translational misalignment, so the scores reflect only how well the 3D volumes match in terms of shape. The translation-invariant versions of these metrics take the max score over a sliding window, defined as:
        \begin{align} \label{eqn:IoU_inv}
            IoU^* & = \max_{i,j,k \in [-n,n]} ~\frac{\sum (\hat{V}_{ijk} \cap V)}{ \sum (\hat{V}_{ijk} \cup V)} \\ 
            CS^* & = \max_{i,j,k \in [-n,n]} ~\frac{<\hat{V}_{ijk}, V>}{ \| \hat{V}_{ijk} \|_2 \| V \|_2 } \,, \label{eqn:CS_inv}
        \end{align}
        where subscripts $i$, $j$, and $k$ represent translations of $\hat{V}$ in each of the three spatial dimensions, bounded by $\pm n$.

    \subsection{Results} 
\begin{figure*}
            \centering
            \includegraphics[width=0.98\textwidth]{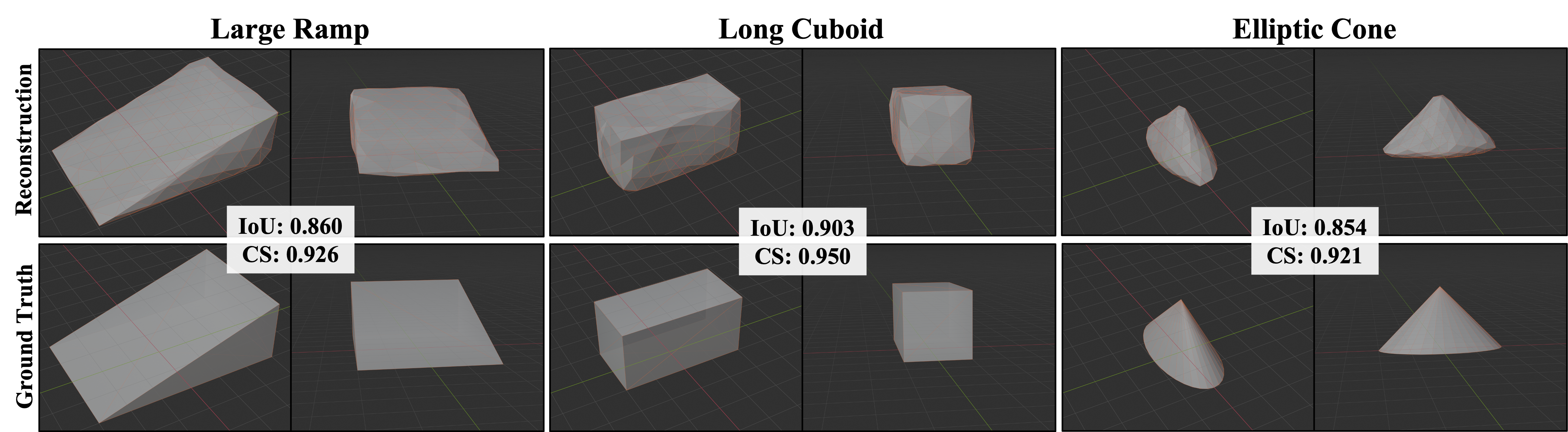}
            \caption{Meshes from median-scoring trial of 3D reconstruction for Large Ramp (train), Long Cuboid (validation), and Elliptic Cone (test).}
            \label{fig:reconstructionResults}
        \end{figure*}

        The experiment is performed on three objects: Large Ramp, Long Cuboid, and Elliptic Cone. The primary object of interest to test is the Elliptic Cone because it was reserved exclusively for testing. Results on the Long Cuboid (validation object) and Large Ramp (training object) serve as points of reference to understand how much the mesh-fitting performance may be impacted by hyperparameter over-tuning and Neural Shader overfitting, respectively. 
        The experiment is run five times for each object at different random seeds. 
        The reconstructed meshes for each object shown in Figure~\ref{fig:reconstructionResults} are sourced from the median-scoring trial, while Table~\ref{table:reconMetrics} provides the mean across all trials for each reconstruction metric.

        The reconstruction scores from Table \ref{table:reconMetrics} trend similarly to what was observed with the $L_{L2}^*$ metric that compared rendering accuracy in Table \ref{table:averageCGANLosses}.
        When comparing Tables \ref{table:averageCGANLosses} and \ref{table:reconMetrics}, it is worth noting that the $L_{L2}^*$ score for the Large Ramp by itself was -10.15 dB, which is better than the average amongst all training objects (-8.93 dB).
        The most noteworthy difference is that the issue of hyperparameter over-tuning appears to be slightly more prominent in the reconstruction results than it is in the rendering accuracy results.         
        This change may be attributable to the hyperparameters introduced by the 3D reconstruction optimization process, which presents additional opportunities for over-tuning.
        Similar to the conclusion from subsection \ref{sec:NeuralShader}.\ref{sec:NeuralShaderResults}, 
        the reconstruction results suggest that the issues of overfitting and over-tuning exist but are both relatively minor.
        
        \setlength{\tabcolsep}{6pt}
        \begin{table}[t] %\begin{wraptable}{R}{0.57\textwidth}
            \centering
            \caption{Mean values of 3D reconstruction metrics across five stochastic trials of the reconstruction experiment.}
            \label{table:reconMetrics}
            \begin{tabular}{|c|c|c|c|c|} 
                \hline %\noalign{\smallskip}
                \textbf{Object} & $\boldsymbol{mIoU}$ & $\boldsymbol{mIoU^*}$ & $\boldsymbol{mCS}$ & $\boldsymbol{mCS^*}$\\
                %noalign{\smallskip}
                \specialrule{.1em}{.05em}{.05em} %special thicker line
                %\noalign{\smallskip}
                Large Ramp & 0.862 & 0.869 & 0.927 & 0.932 \\ 
                \hline
                Long Cuboid & 0.902 & 0.902 & 0.949 & 0.949 \\ 
                \hline
                Elliptic Cone & 0.853 & 0.853 & 0.921 & 0.921 \\ 
                \hline
            \end{tabular}%}
        \end{table}%\end{wraptable}
        \setlength{\tabcolsep}{1.4pt}
        
\section{DISCUSSION}\label{sec:Discussion}
    The method proposed in this work has been successful as a proof of principle. However, there are areas for improvement that should be addressed in the future to help this become a more pragmatic tool for real-world use.
    This includes support for multiple materials and object segments, support for multistatic modalities, and additional validation on real world data.
    There are also open research problems related to differentiable rendering for SAR that should be the subject of future work. These include additional applications and an alternative approach based on SAR raw signal simulators.

    \subsection{Future Improvements}
        In the method's current form, objects are assumed to be of a uniform material, and changing the assumed material requires new corresponding weights for the Neural Shader. Support for multiple material types, as well as backpropagation to material properties, would  broaden the tool's utility to other differentiable SAR rendering applications. This could manifest itself as additional feature maps output by the Feature Rasterizer to convey material property context to the Neural Shader.
        
        The current method also only supports one object segment (i.e., one mesh) at a time. Support for multiple segments would allow handling of background clutter, which is typically present in real SAR images, and may make support for multiple materials easier to implement.
        
        This work also only considered monostatic SAR collections. Generalizing the approach to accommodate multistatic SAR modalities (where transmitter and receiver are \textit{not} co-located) would expand the coverage of SAR collection modes. 

        Additionally, since this differentiable rendering method is based on rasterization, it is limited to modeling first-order (single-bounce) effects. 
        Depending on object geometry and material, SAR images may have pronounced multi-path effects as well, which cannot be fully captured as-is. 
        In our internal experiments, there were not significant differences between versions of the data simulated with single vs. double-bounce effects, but other settings could differ.
        A differentiable rendering scheme based on raytracing may be able to capture such multi-path effects, but is outside the scope of this current work.

\begin{figure*}[t]
            \centering
            \includegraphics[width=0.96\textwidth]{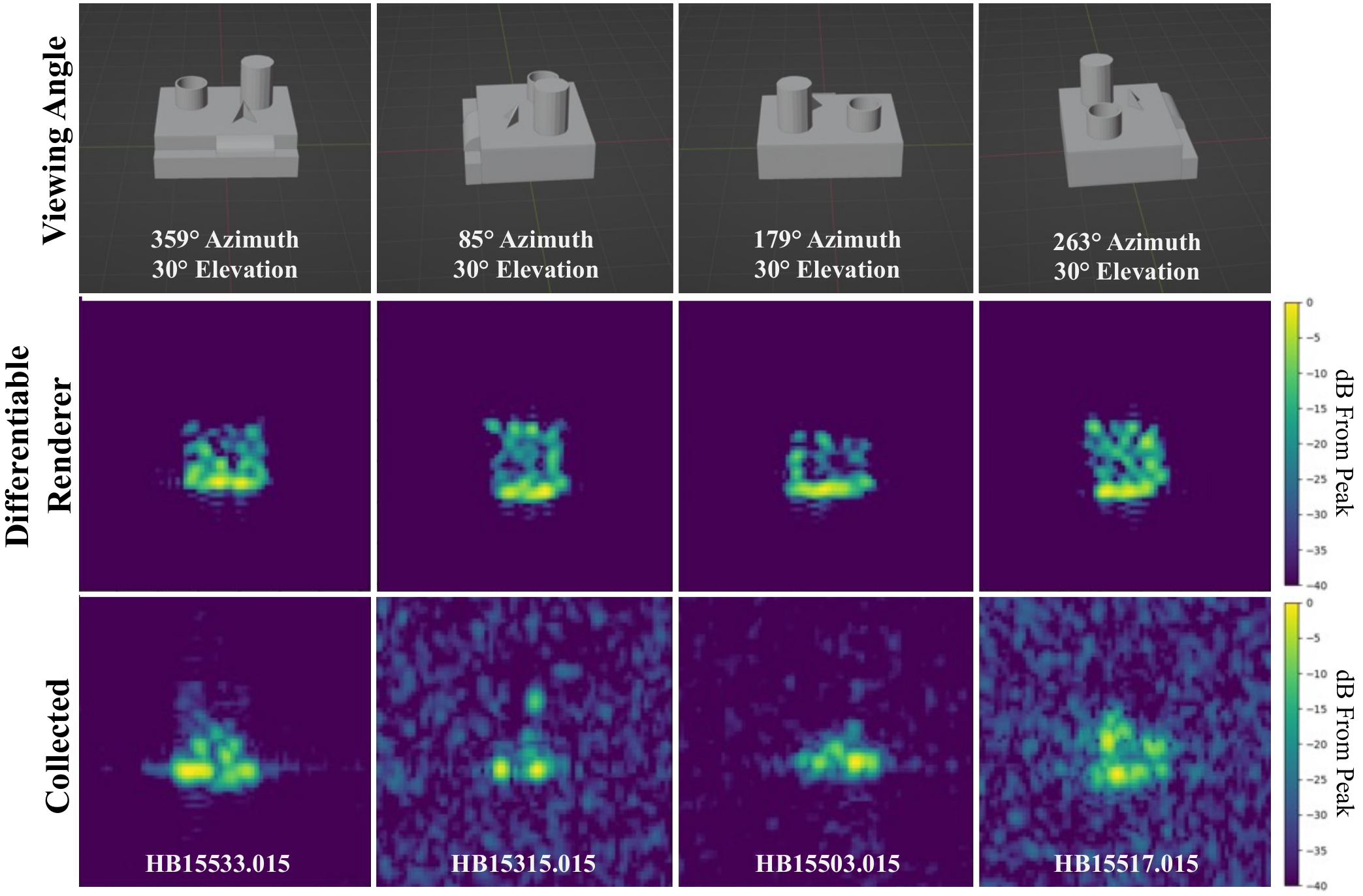}
            \caption{Example images from our differentiable renderer compared to collected images of the SLICY object.}
            \label{fig:comparisonToSlicy}
        \end{figure*}

        In real-world SAR systems, contributions from facets that are equidistant from the sensor are coherently summed, causing the effect known as layover.
        However, in the differentiable renderer's current form, contributions from facets that get mixed due to layover are combined (non-coherently) via a weighted mean. 
        This is a limitation shared with \cite{SarViz06}, which utilizes alpha blending. 
        This may partially explain why the peaks are not as sharp on images produced by the differentiable renderer where layover is known to occur, like in Figure \ref{fig:comparisonToSlicy}.
        
        Lastly, the SAR data used to train and evaluate our method was synthetically generated. 
        The cost of collecting large quantities of measured data make it prohibitively expensive to use for training.
        However, future experiments should at least use measured data for validation and test objects.
    
    \subsection{Future Directions} 
        We chose the task of 3D object reconstruction for demonstration purposes, but future work should investigate other applications enabled by differentiable rendering.
        Particularly, we would like to see this applied to the task of generating adversarial examples subject to 3D shape or material constraints. This task may be useful on its own for assisting in vehicle design or augmentation, but also enables the downstream tasks of geometrically-constrained data augmentation and adversarial training for deep neural networks.
        
        An approach based on SAR raw signal simulation would be fully coherent and could capture effects from transmitter and receiver polarizations.
        It would also allow the user to employ an image formation algorithm of choice. Additionally, it would eliminate the need to train a CGAN for shading, which is cumbersome because of the large number of high-fidelity SAR images needed to train it. 
        However, such an approach would likely be computationally more costly than the one proposed in this paper, which may hinder its utility in optimization problems. 
        
    \subsection{Comparison to Collected Data}
        This subsection provides a visual comparison of imagery produced by the differentiable SAR renderer to collected imagery at comparable geometries. Although the differentiable renderer is not intended for use as a general SAR simulator, comparison to additional data sources is helpful for understanding its capabilities and limitations.
        
        Publicly available SAR data with corresponding scene representation ground-truth information (i.e., detailed CAD models) is difficult to come by, and prohibitively expensive to collect.
        Fortunately, the Moving and Stationary Target Acquisition and Recognition (MSTAR) public dataset \cite{Mstar98} includes a synthetic calibration target known as SLICY (Sandia Laboratory Implementation of Cylinders). 
        While detailed CAD models are not directly available for SLICY, information about the size and shape of the components that compose it is known \cite{SarImgGenOfSlicy10, ModelGenFromSarImgs01}, as well as general information about its material (i.e., smooth metal).
        We manually constructed a CAD model that is nominally matched to the geometry of SLICY, seen in Figure \ref{fig:comparisonToSlicy}. 
        
        The SLICY data was collected at X-band, but at a resolution about 4x as coarse as what was used in our main experiments, and image-formation was done in the slant-plane.
        We resampled the SLICY imagery to project it into the ground-plane and change its pixel-spacing to 0.15m in both range and cross-range.
        
        To make images from our differentiable renderer comparable, we down-sampled the images it produced to match the resampled SLICY data described above. 
        Also, since the differentiable renderer directly produces images with a PEDF-remapping, we (approximately) reverse-processed the remap to estimate what the raw pixel magnitudes would be. 
        Since the PEDF-remap is lossy, the reverse-processing is imperfect; most notably, only roughly 30dB of dynamic range is retained due to clipping that occurs as part of PEDF-remapping.

        Lastly, because scaling differs between image sources, we must normalize each image in order to compare them on a common scale.
        We normalize each image by its peak magnitude, and display the images on a decibel scale with 40dB of dynamic range.
        The resulting images are displayed in Figure \ref{fig:comparisonToSlicy}.

        Despite a number of mismatches impacting the comparison in Figure \ref{fig:comparisonToSlicy}, the images maintain good structural similarity. 
        The most prevalent difference seems to stem from the differing material types, which explains why a greater proportion of the object's body appears illuminated in the rendered images compared to the collected counterparts.
        The SLICY object was also explicitly designed to demonstrate certain multi-path effects, which the differentiable renderer does not emulate. 
        The hollow short cylinder is one such example, and it seems to be the source of the top-right scatterer in collected image HB15315.015. The delayed return from the cavity is what causes the scatterer to appear displaced in range.

\section{CONCLUSION}\label{sec:Conclusion}
    In this work, we described an approach for differentiable rendering of SAR imagery which is a composition of techniques from computer graphics and neural rendering. We presented proof-of-concept results on the task of 3D object reconstruction and discussed the approach's current limitations. To the best of our knowledge, this is the first successful demonstration of differentiable rendering for SAR-domain imagery, which we hope will serve as a useful starting point for others seeking to use differentiable SAR rendering in their research.

\section*{ACKNOWLEDGMENT}\label{sec:Acknowledgements}
    The authors would like to thank the editors and anonymous reviewers for their helpful feedback during the review process, along with KBR inc. for partial support of the project. 
    The authors would also like to thank Emeritus Professor Hao Ling for his helpful advice at the project’s onset, as well as Nonie Arora for her help in reviewing the manuscript. 

\bibliographystyle{IEEEtran}
\bibliography{SarRendering}

\end{document}